# Electromagnetic Absorption as a Novel Tool in Chemical Vapor Deposition Toolbox: Ultra-Fast Growth of Continuous Graphene Film

Hadi Arjmandi-Tash and Grégory F. Schneider[†]

*Leiden University, Faculty of Science, Leiden Institute of Chemistry, Einsteinweg 55, 2333CC Leiden, The Netherlands*

[†] to whom correspondence should be addressed: g.f.schneider@chem.leidenuniv.nl

Abstract

This report introduces the electromagnetic absorption of the catalyst as a novel knob in optimizing chemical vapor deposition of graphene. A chromium film initially deposited at one side of a catalyst copper foil absorbs radiative thermal energy several times higher than the plain copper foil and migrates to the other side of the foil in the form of microscale grains. The process involves local melting of the copper which together with the increased operation temperature, improves the crystalline quality of growing graphene. Local melting of copper replaces the otherwise-necessary annealing process; dramatically lowering the process duration and costs. Now a continuous graphene film can be grown in only five minutes, nominating the protocol as the shortest ever reported. The observations reported in this manuscript boost the general knowledge about metal mixture, additional to promoting graphene growth science and technology.

Introduction

Since the introduction of chemical vapor deposition as an efficient approach to synthesize graphene [1], the process has been optimized/adjusted to improve/customize the properties of graphene [2], [3]. Increasing the crystalline quality [4], [5] and global uniformity of graphene [6], [7] as well as improving the throughput [8], [9] and minimizing the production cost [10], [11] along with the generalization of the process [12]–[15] are among the targets. Traditionally, copper serves as an important catalyst for the growth of graphene. Thanks to the low carbon solubility of carbon in copper, the growth is self-limited to produce (primarily) monolayer graphene. Interestingly, the chemical vapor deposition (CVD) process temperature − in annealing and growth phases − approaches the melting point of the copper ($T_m^{Cu}$). Though it is technically complex, graphene can be grown on molten copper also [16]. The number of the crystalline defects including grain boundaries − as potential graphene nucleation sites − in molten copper is limited which can eventually lead to crystalline monolayer graphene [17]. Indeed by using appropriate supporting layers to prevent the de-wetting of the copper foil, epitaxial growth of single crystal domains of ~200 μm has been realized above $T_m^{Cu}$ [16]. As the second advantage of using molten copper, the immobilized catalytic species at the surface lowers the crystalline defects (e.g. voids) in growing graphene. Though relevant experimental evidence can be hardly found in literature, it has been shown theoretically that high mobility and diffusion of the surface catalyst atoms drives "defect healing" processes in which defects such as pentagons and heptagons promote to carbon hexagons [18]. We target this aspect in this report.

Mixing (alloying) metals enable to combine the favorable properties of different metals for specific applications; this remains yet a seldomly touched strategy in the growth of graphene. As one of the few successful reports, nickel and molybdenum were rationally combined to achieve self-limiting growth with outstanding reproducibility [19]. Here, the precipitated carbon species form strong and stable bonds with molybdenum and are excluded from the growth to yield strictly single-layer graphene. Separately the catalytic capability of the copper was dramatically improved by alloying with nickel [20]; Indeed the nickel-mediated segregation of carbon radicals in the Cu-Ni alloy system can boost the growth rate by an order of magnitude over singular copper.

In this paper, we report a peculiar approach in growing graphene on a bi-component substrate composed of copper and chromium: chromium species, initially deposited at the backside of the copper foil migrate thought the bulk copper to the other side, where graphene is to be grown. Copper serves as the traditional catalyst to mediate cracking of methane molecules into carbon radicals. Dilute chromium species on the other hand, locally melt the copper and run the defect healing mechanism to improve the crystalline quality of forming graphene lattice.

CVD of graphene is traditionally performed in tube oven (= "hot-wall") chambers [7], [21], [22] with the heating element placed outside the chamber. Symmetrical thermal radiation forms a uniform thermal zone in which the specimen (e.g. copper foil) receives a homogenous heat flux (= flow of thermal energy per unit area and time) from surrounding. The heating in "cold-wall chambers" [10], [23]–[25] is heterogeneous: Here a heating stage is placed inside the chamber which directly heats-up the specimen in contact, via thermal conduction. The radiated thermal energy, on the other hand, exposes one side of the copper foil. The camber features a huge thermal gradient between the stage (normally at T > 1000 °C) and the walls (normally at T ~ 100 °C) which eventually provide a non-uniform heating zone which potentially can lead to random growth. Although graphene growth in cold-wall chambers is cost-effective, the heterogeneous heating is a major drawback. In this letter, however, we benefit from this heterogeneity to drive chromium specious from one side to the other side of the copper foil during the growth of graphene. We demonstrate that our approach − uniquely feasible in the cold-wall chambers − considerably boosts the crystalline quality of graphene while minimizing the growth time and costs.

Experiments

The test sample is a copper foils (Alfa Aesar, 99.999% purity, 25 μm thickness) on which a chromium film of 50 nm was evaporated. We placed the sample on the hot stage of a cold-wall CVD set-up (nanoCVD-8G, Moorfield Nanotechnology) with the chromium deposited side facing the stage (we will refer to this side as the "bottom side" throughout this manuscript, see the schematic in inset Figure 1-a). In an earlier publication [11], we characterized and compared graphene grown in a "cold-wall" and "hybrid cold/hot wall" chambers, concluding the graphene grown via the conventional cold-wall approach is less uniform and of poorer quality. We adopted this unsuccessful recipe – including subsequent annealing and growth phases – to grow graphene in this work (see the Methods). We varied the growth temperature and the annealing duration but used a fixed growth duration of three minutes in different experiments. Later on, we will show that with an annealing duration of seven minutes at 1035°C, chromium migrate to the top side where graphene grows during the growth phase. Figure 1-a displays the top side of a copper foil initially and partially covered with chromium film at the bottom left side (see the inset schematic). The right side of the sample (with no chromium film deposited) appears shiny and smooth; the migration of the chromium film to the top side, however, turns the sample matte (rough) on the left side. High resolution atomic force microscopy (AFM) and scanning electron microscopy (SEM, Figure 1-b,c and d) evidence that the diffusion of the chromium have induced a complex microstructure at the surface of the foil: The surface has split into a landscape of two phases where a lath structure with sharp edges pops-out from the background mattress (Figure 1-b and c). Standard copper etching solution (ammonium persulfate) dissolves the background while the lath structure is well dissolved in chromium etcher (supplementary information). The experiment implies that copper remains the major constituent of the background while the lath structure mainly consists of chromium (although a trace amounts of one element in another is possible).

We characterize the quality of the grown graphene by means of Raman spectroscopy (Figure 1-e). In agreement with our earlier publication [11], the growth recipe provides poor quality graphene (or even amorphous graphite) on plain copper side. The graphene grown on the Cu/Cr side, however, exhibits standard Raman peaks (G peak at ~1580 cm$^{-1}$ and 2D peak at ~2680 cm$^{-1}$) and is of superior crystalline quality, evidenced by a negligible D peak (at 1350 cm$^{-1}$).

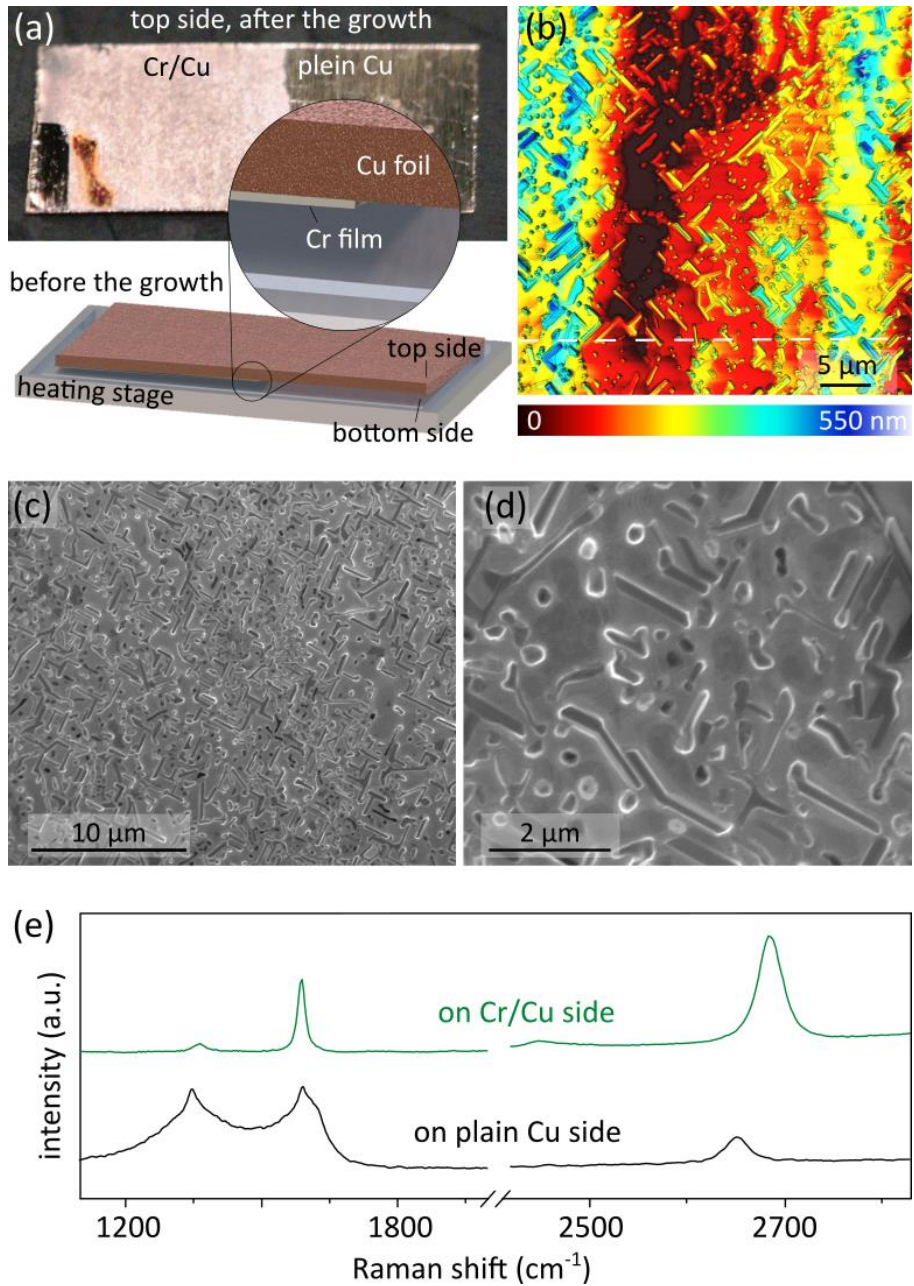

Figure 1: Graphene grown on Cr/Cu structure
a) Optical micrograph of a test sample (2 cm × 1 cm) after the growth of graphene: the bottom side of the copper foil (facing the heating stage) was initially covered with a thin (~100 nm) chromium film (demonstrated in the inset). During the growth, chromium drives to and makes the top surface rough.
b) Representative atomic force micrograph featuring chromium texture popped-out from the copper background on left side of the sample in a.
c) and d) Representative low- and high-resolution scanning electron micrographs of the left side of the sample in a.
e) Comparison of the Raman spectroscopy data recorded on the left (on Cr/Cu) and right (plain Cu) sides in a.

Figure 2Figure 2 provides an in-depth Raman characterization of the sample after the growth of graphene. The migration of the chromium from the bottom to the top side is evident in the optical micrograph in a. We identified low-frequency Raman spectral bands which are sensitive to the chemical constituents of the substrate. The chromium phase manifests itself as a strong peak centered at 85 cm$^{-1}$ and is distinct from the background mattress with a low-amplitude signature at its right shoulder (inset Figure 2-a, b and c). Mappings of the Raman characteristic peaks of graphene are provided in panels d to f. The narrow 2D peak (FWHM < 50, panel d) and large $I_{2D}/I_G$ ratio (> 1, panel e) are the signatures of a monolayer graphene [26]. The graphene is of pronounced crystalline quality evidenced by a negligible $I_D/I_G$ ratio (panel f). Interestingly, the properties of the graphene is independent of the local structure (texture) of the underlying substrate as no correlation between the mappings in d-f and b are observed. This is an important evidence to explain the migration of the chromium and the improved quality of the graphene which will be discussed later.

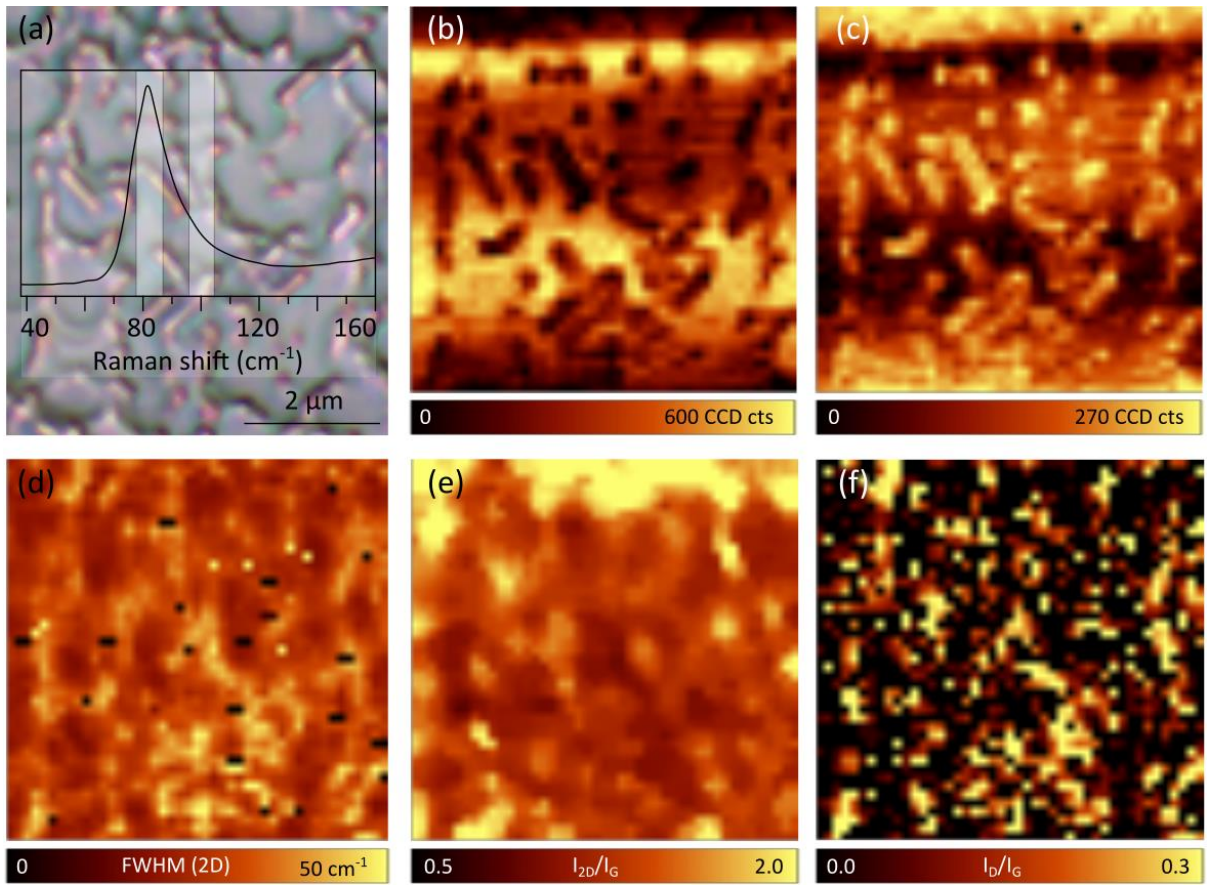

Figure 2: Raman characterization of graphene on the Cr/Cu structure
a) Optical micrograph of a selected window featuring chromium structures migrated to the front side of the Cu foil after ten minutes of growth (seven minutes of annealing) at 1035 °C. Inset shows a low frequency peak sensitive to Cr/Cu composition in Raman spectrum.
b) Mapping the amplitude of the Raman signal at the narrow band centered at 100 cm$^{-1}$ (see the spectrum in a). The band corresponds to the chromium microstructures.
c) Mapping the amplitude of the Raman signal at the narrow band centered at 85 cm$^{-1}$ (see the spectrum in a). The band corresponds to the copper background.
d) Mapping the width of the graphene Raman 2D peak (centered at ~2670 cm-1): The width of the 2D peak hardly exceeds 50 cm$^{-1}$, manifesting that the graphene is predominantly monolayer.
e) Mapping the relative intensity of the 2D ($I_{2D}$, centered at ~2680 cm$^{-1}$) and G ($I_G$, centered at ~1580 cm$^{-1}$) peaks: The $I_{2D}/I_G$ ratio stays mainly above one as another indication of monolayer graphene.
f) Mapping the relative intensity of the D ($I_D$, centered at ~1350 cm$^{-1}$) and G ($I_G$, centered at ~1580 cm$^{-1}$) peaks: The $I_D/I_G$ ratio stays close to zero indicating a negligible amount of the crystalline defects.

The migration of the chromium from the bottom to the top side of the copper starts during the annealing phase of the growth. Figure 3-a correlates the Raman spectra and the surface properties of several graphene samples grown at 1035°C with varied annealing durations. The growth duration (after the annealing) is set on three minutes for all the experiments. Annealing durations below six minutes (i.e. the total process duration below nine minutes) have negligible effect on the surface corrugations on the foil and are insufficient to have chromium migrated to the top side. Here, the growth process is similar to conventional approaches with an insufficient (short) copper annealing phase leading to a poor graphene quality, manifested by a considerably large amplitude D peak. Chromium traces starts to appear in the sample with the total process duration of ten minutes (referred to as optimize process duration, $t^*$). Graphene quality is the highest in a tight "temporal window" ($\Delta t^* \sim 1$ minute at 1035°C) close to $t^*$. Longer process durations, however degrades graphene.

The optimized process duration depends on how fast chromium migrates to the top side of the copper foil which itself is a strong function of the growth temperature. Figure 3-b plot the $t^*$ at various growth temperatures depicting a linear correlation between 1005°C and 1045°C. Interestingly, while a minimum total process duration of $t^*= 20$ (with $\Delta t^*= 6$ min) minimizes the crystalline defects at 1005°C, the process could be as short as five minutes (two minutes of annealing followed by three minutes of growth) to achieve a decent graphene quality at 1045°C. The allowed temporal window is tighter at elevated temperatures, demanding an efficient temperature control unit. Increasing the growth temperature above 1045°C does not affect $t^*$ and $\Delta t^*$ considerably.

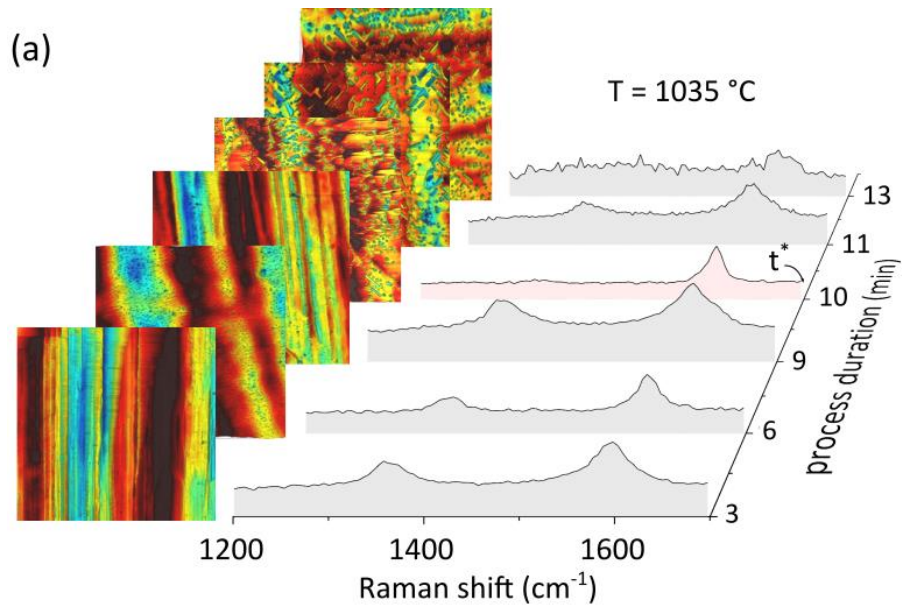
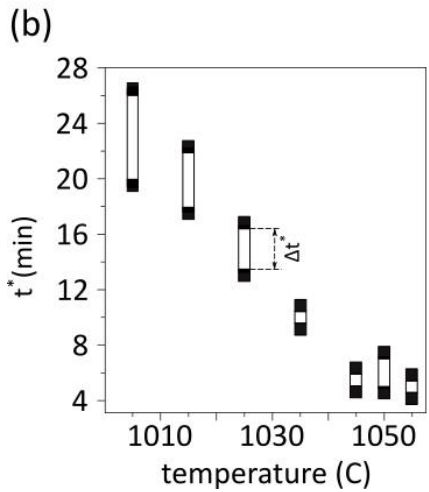
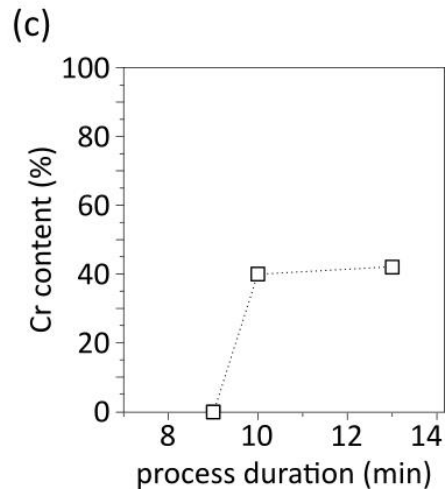

Figure 3: Quality of the graphene as a function of the process (annealing) duration

a) Raman spectra of several graphene samples grown with different process durations, ranging from three to thirteen minutes (annealing duration ranging from zero to ten minutes); $t^*$ marks the optimum process duration which provides the lowest D peak (highest crystalline quality). The insets on left shows the AFM characterization of the surface of the corresponding sample.

b) Optimum process duration ($t^*$) of different samples as a function of the growth temperature. $\Delta t^*$ (open rectangular markers) corresponds to the temporal window over which graphene quality persists the highest. Filled markers correspond to uncertain regions in between subsequent samplings.

c) Chromium content as a function of the process duration estimated by the X-ray photoelectron spectroscopy (XPS) at the top surface of the foil

Few mechanisms might be considered in action to drive the chromium film from one side to the other side of the copper foil: Mixing the chromium and copper to form a conventional binary alloy is a potential scenario. The operation temperature of the CVD set-up remains below the solidus line in Cr-Cu phase diagram (1076 °C [27]) illustrating that the conventional alloying process – including mixing molten components – is irrelevant here. Random thermal motion of the materials formulated by the Fick's law of diffusion[28] (including grain boundary and lattice diffusion[29]), on the other hand, should be independent of the placement of the sample with respect to the heating stage and would eventually result in a uniform concentration across the foil. The chromium film in our experiments, however, migrates through the copper only when it initially faces the heating stage and is specific to the cold-wall chambers: Indeed similar thermal processes with flipped samples (chromium film initially facing up) or inside a tube-oven chamber (instead of a cold-wall chamber) did not cause the migration of the chromium (Supplementary Information). X-ray photoelectron spectroscopy (XPS) analysis of the samples, furthermore, revealed the concentration of the chromium in the top copper side exceeds 40% (Supplementary Information); indeed the governing mechanism in our system, shifts a high concentrated chromium region from one side to the other side of the foil. Thermal diffusion of solids in which a temperature gradient energizes a material of a certain thermodiffusion coefficient to diffuse, might explain the directionality of the observation: in-contrast to the tube-oven chambers which provide a uniform heating zone, there is a huge thermal gradient between the heating stage (T > 1000 °C) and the walls (~ 100 °C) in cold-wall reaction chambers. A naive hypothesis may argue that the foil "feels" this thermal gradient, causing the chromium atoms to diffuse from the hotter side (facing the heater) to the colder side (facing the cold wall, in-contact with the fresh operation gases) of the copper foil, but not in reverse direction. Our finite element simulations, however, rules out this scenario as the high thermal conductivity of copper results in a negligible temperature gradient in between its faces (Supplementary Information).

Any successful scenario explaining our observations has to be built-up based on two important facts. Firstly, at the elevated operation temperature, the heating stage acts as a (semi-) black body radiator, emitting electromagnetic waves in near-infrared spectrum (inset Figure 4-a). Copper is a reflective material in this spectral range exhibiting a negligible absorption of 7% at elevated temperatures [30]. In contrary, chromium is known as a lossy metal with electromagnetic absorption of ~40% [31]. In other words, while uncoated copper reflects back some 93% of upcoming radiation, chromium coating dramatically improves radiative heat transferring by almost six folds. Thermal conduction remains yet a parallel heat transfer mechanism. We modeled the heat flow inside the oven considering thermal convection (by the process gases) and radiation to the cold-walls (Figure 4-a, see SI for the details of modeling). Interestingly, by powering-up the heater, the temperature of the uncoated copper foil falls below that of the stage by ~100°C. Note that the quality of chemically synthesized graphene is highly sensitive to the reaction temperature as insufficient heating fails to provide the necessary activation energy to decompose the precursors [32]. Indeed this temperature difference between the foil and the stage explains the typically seen poor quality of graphene in cold-wall chambers [7], [11]. The improved thermal energy absorption, however, drives the chromium coated copper foil to follow the temperature of the stage closely. The higher growth temperature, however, is not the only origin of the increased graphene quality in chromium coated copper foil.

As another important fact, certain observations demonstrate that the migration of the chromium is accompanied by locally melting the copper foil. A first evidence appears by comparing the morphology of the foils (top surface) with and without chromium (Figure 4-b): The plain copper side features gradual undulations of the surface with the amplitude and wavelength of few hundreds of nanometers; the wavy surface is a result of the fabrication process of the copper foil and is typically observed in graphene growth studies on commercial copper foils [34]. Migrated chromium, on the other hand, form large mesas of ~ 80 nm. Interestingly, the surface of the copper in between the chromium mesas has been flattened which would be possible only after a melting process. Additionally, we regularly observed grooves forming between the Cr/Cu and plain Cu sides (Figure 4-c), indicating the presence of a liquid phase during the operation: indeed the transformation of a melt (lower density thus higher volume) to a solid (higher density thus lower volume) comes along with a shrinkage in the volume causing the groove [35]. It worth noting that similar grooves are typically seen at the melt/mold interface after solidifying a molten metal in a casting process [36]. Chromium assisted melting of the copper is the major cause of the improvement in graphene crystalline quality: In fact the higher mobility and longer surface diffusion range of the molten copper atoms drive a "defect healing" processes in which the structural defects (e.g. pentagons and heptagons) transform into perfect hexagonal rings, as observed by quantum chemical molecular dynamics simulations [18]. Experimentally, defect-free graphene achieved by chemical vapor deposition has already been demonstrated on molten copper foil [16], [17].

We now propose a potential mechanism to explain our observations (Figure 4-d): Chromium deposited on the bottom side of the copper absorbs electromagnetic radiation – well beyond the plain copper – and starts melting neighboring layers of the copper foil. Melting the copper opens up paths for the migration of the chromium. In an optimize timing, the growth of the graphene would start when the most top copper layers are in molten state but chromium has not reach there yet. The defect healing process on molten copper foil achieves the highest graphene quality. In a delayed process, the presence of the chromium on the top layer repels the catalyst copper and degrades the graphene. Presence of molten copper layers in-between the chromium and forming graphene justifies the lack of any correlation between the Raman mappings of graphene and the underlying features on the foil (Figure 2). The rate of advancing the chromium front scales inversely with the temperature which matches with the longer $t^*$ measured at lower temperatures (Figure 3-bFigure 2). Once the molten layer forms at the top copper surface, the optimized growth occurs in a larger temporal window (larger $\Delta t^*$) at lower temperatures before chromium reaching the top surface.

We note that the improved radiation absorption by the chromium cannot directly cause the melting of the copper foil as the temperature of the heating stage remains below the melting point of copper. Particularly we noticed the migration of the chromium at temperatures as low as 870°C (~ 200°C below the melting point of copper). Surface melting phenomenon [18], [35] in which a thin layer (typically of few nanometer thickness) at the surface melts below the melting point of the bulk is an important consideration. In-fact the reduced coordination degree of atoms accounts for this advance melting. The phenomenon lowered the melting temperature at the surface of lead by 25 %[35]. The same phenomenon might be on-going at the Cu/Cr interface in our work. Our findings, however, are insufficient to validate (or rule-out) this hypothesis and more systematic investigations must be followed-up.

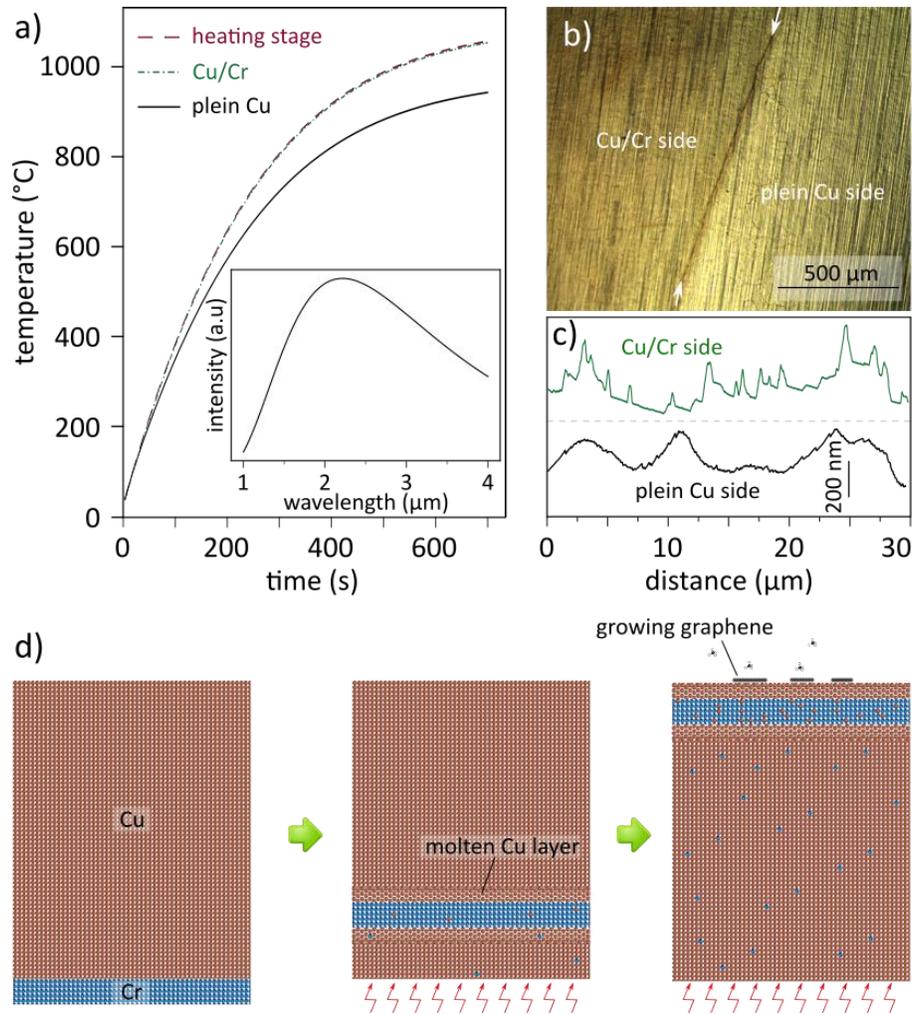

Figure 4: A potential scenario explaining the migration of the chromium film accompanied by an improved graphene quality

a) Simulation of the temperature of the heating stage, chromium-coated and plain copper foil upon powering up the resistive heater; inset: black-body radiation spectrum at 1035 °C, formulated by the Planck's equation [38]

b) Optical micrograph focusing at the border of the Cr/Cu and plain Cu sides after ten minutes of the process at 1035°C. A groove appears and splits the two sides, marked by white arrows.

c) Representative surface profiles over the Cr/Cu (corresponding to the dash-line in Figure 1-b) and plain Cu sides

d) A potential mechanism explaining the migration of the chromium from the bottom to the top sides of the copper foil: The chromium film absorbs electromagnetic energy radiated by the hot stage at the elevated temperatures and partially melts the neighboring layers of the copper foil. The chromium film moves up through the molten copper; with an appropriate timing, the growth of graphene on the molten copper layers improves the crystalline quality of graphene.

## Conclusion

We introduce a peculiar approach in chemical vapor deposition of graphene in which the migration of a chromium film, initially deposited on one side of a copper foil to the other side causes the local melting of the foil and eventually boosts the quality of the graphene. A continuous graphene sheet now is grown in less than five minutes as a result of eliminating the surface activating the copper foil by thermal annealing. We investigate the mechanism driving the chromium species through the copper. Our report nominates electromagnetic absorption as a novel tool in the CVD toolbox to further improve the efficiency of the approach, being of great potential in scalable and cost-effective preparation of high-quality graphene. The findings in this report promote the science of interfacing metals.

## Acknowledgements

The work leading to this work has gratefully received funding from the European Research Council under the European Union's Seventh Framework Programme (FP/2007-2013)/ERC Grant Agreement No. 335879 project acronym 'Biographene', and the Netherlands Organisation for Scientific Research (Vidi 723.013.007).

## References

[1] X. Li *et al.*, "Large-area synthesis of high-quality and uniform graphene films on copper foils.," *Science*, vol. 324, no. 5932, pp. 1312–1314, 2009.

[2] C. Mattevi, H. Kim, and M. Chhowalla, "A review of chemical vapour deposition of graphene on copper," *J. Mater. Chem.*, vol. 21, no. 10, pp. 3324–3334, Feb. 2011.

[3] W. Choi, I. Lahiri, R. Seelaboyina, and Y. S. Kang, "Synthesis of Graphene and Its Applications: A Review," *Crit. Rev. Solid State Mater. Sci.*, vol. 35, no. 1, pp. 52–71, Feb. 2010.

[4] Y. Hao *et al.*, "The Role of Surface Oxygen in the Growth of Large Single-Crystal Graphene on Copper," *Science (80-. )*, vol. 342, no. 6159, pp. 720–723, Nov. 2013.

[5] G. H. Han *et al.*, "Influence of Copper Morphology in Forming Nucleation Seeds for Graphene

Growth," *Nano Lett.*, vol. 11, no. 10, pp. 4144–4148, Oct. 2011.

[6] Q. Yu *et al.*, "Control and characterization of individual grains and grain boundaries in graphene grown by chemical vapour deposition," *Nat. Mater.*, vol. 10, no. 6, pp. 443–449, Jun. 2011.

[7] Z. Han *et al.*, "Homogeneous optical and electronic properties of graphene due to the suppression of multilayer patches during CVD on copper foils," *Adv. Funct. Mater.*, vol. 24, no. 7, pp. 964–970, 2014.

[8] S. Bae *et al.*, "Roll-to-roll production of 30-inch graphene films for transparent electrodes.," *Nat. Nanotechnol.*, vol. 5, no. 8, pp. 574–8, Aug. 2010.

[9] J.-H. Lee *et al.*, "Wafer-Scale Growth of Single-Crystal Monolayer Graphene on Reusable Hydrogen-Terminated Germanium," *Science (80-. ).*, no. April, pp. 1–6, 2014.

[10] T. H. Bointon, M. D. Barnes, S. Russo, and M. F. Craciun, "High Quality Monolayer Graphene Synthesized by Resistive Heating Cold Wall Chemical Vapor Deposition," *Adv. Mater.*, vol. 27, no. 28, pp. 4200–4206, 2015.

[11] H. Arjmandi-Tash, N. Lebedev, P. van Deursen, J. Aarts, and G. F. Schneider, "Hybrid cold and hot-wall chamber for fast synthesis of uniform graphene," *Carbon N. Y.*, vol. 118, pp. 438–442, 2017.

[12] W. Yang *et al.*, "Epitaxial growth of single-domain graphene on hexagonal boron nitride," *Nat. Mater.*, no. July, pp. 792–797, 2013.

[13] S. Tang *et al.*, "Silane-catalysed fast growth of large single-crystalline graphene on hexagonal boron nitride.," *Nat. Commun.*, vol. 6, p. 6499, 2015.

[14] H. Arjmandi-Tash *et al.*, "Large scale graphene/h-BN heterostructures obtained by direct CVD growth of graphene using high-yield proximity-catalytic process," *J. Phys. Mater.*, vol. 1, no. 1, p. 015003, Sep. 2018.

[15] H. Arjmandi-tash, "In situ growth of graphene on hexagonal boron nitride for electronic transport applications," *arXiv Prepr. arXiv 1701.06062*, 2017.

[16] Y. A. Wu *et al.*, "Large Single Crystals of Graphene on Melted Copper Using Chemical Vapor

Deposition," *ACS Nano*, vol. 6, no. 6, pp. 5010–5017, 2012.

[17]     D. Geng *et al.*, "Uniform hexagonal graphene flakes and films grown on liquid copper surface," *Proceeding Natl. Acad. Sci. PNAS*, vol. 109, no. 21, pp. 7992–7996, 2012.

[18]     H.-B. Li *et al.*, "Graphene nucleation on a surface-molten copper catalyst: quantum chemical molecular dynamics simulations," *Chem. Sci.*, vol. 5, no. 9, pp. 3493–3500, Jul. 2014.

[19]     B. Dai *et al.*, "Rational design of a binary metal alloy for chemical vapour deposition growth of uniform single-layer graphene," *Nat. Commun.*, vol. 2, no. May, p. 522, 2011.

[20]     T. Wu *et al.*, "Fast growth of inch-sized single-crystalline graphene from a controlled single nucleus on Cu–Ni alloys," *Nat. ma*, vol. 15, pp. 43–47, 2016.

[21]     X. Li *et al.*, "Large-area synthesis of high-quality and uniform graphene films on copper foils SOI," *Science*, vol. 324, no. 5932, pp. 1312–4, Jun. 2009.

[22]     X. Li *et al.*, "Large-Area Graphene Single Crystals Grown by Low-Pressure," *J. Am. Chem. Soc.*, vol. 133, no. 9, pp. 2816–9, 2011.

[23]     A. I. S. Neves *et al.*, "Transparent conductive graphene textile fibers," *Sci. Rep.*, vol. 5, p. 9866, 2015.

[24]     V. Miseikis *et al.*, "Rapid CVD growth of millimetre-sized single crystal graphene using a cold-wall reactor," *2D Mater.*, vol. 2, no. 1, p. 014006, 2015.

[25]     N. Mishra, V. Miseikis, D. Convertino, M. Gemmi, V. Piazza, and C. Coletti, "Rapid and catalyst-free van der Waals epitaxy of graphene on hexagonal boron nitride," *Carbon N. Y.*, vol. 96, pp. 497–502, 2016.

[26]     a C. Ferrari *et al.*, "Raman Spectrum of Graphene and Graphene Layers," *Phys. Rev. Lett.*, vol. 97, no. 18, p. 187401, Oct. 2006.

[27]     K. Zeng and M. Hämäläinen, "Thermodynamic analysis of stable and metastable equilibria in the Cu-Cr system," *Calphad*, vol. 19, no. 1, pp. 93–104, Mar. 1995.

[28]     D. Gupta, Ed., *Diffusion Processes in Advanced Technological Materials*.     Springer-Verlag Berlin Heidelberg, 2005.

[29] R. M. German and R. M. German, *Thermodynamic and Kinetic Treatments*. Butterworth-Heinemann, 2014.

[30] I. Setién-Fernández, T. Echániz, L. González-Fernández, R. B. Pérez-Sáez, and M. J. Tello, "Spectral emissivity of copper and nickel in the mid-infrared range between 250 and 900 °C," *Int. J. Heat Mass Transf.*, vol. 71, pp. 549–554, Apr. 2014.

[31] P. Johnson and R. Christy, "Optical constants of transition metals: Ti, V, Cr, Mn, Fe, Co, Ni, and Pd," *Phys. Rev. B*, vol. 9, no. 12, pp. 5056–5070, Jun. 1974.

[32] Y. Zhang, L. Zhang, and C. Zhou, "Review of chemical vapor deposition of graphene and related applications," *Acc. Chem. Res.*, vol. 46, pp. 2329–2339, 2013.

[33] P. Procházka *et al.*, "Ultrasmooth metallic foils for growth of high quality graphene by chemical vapor deposition," *Nanotechnology*, vol. 25, no. 18, p. 185601, May 2014.

[34] A. J. Marsden, M. Phillips, and N. R. Wilson, "Friction force microscopy: a simple technique for identifying graphene on rough substrates and mapping the orientation of graphene grains on copper," *Nanotechnology*, vol. 24, no. 25, p. 255704, Jun. 2013.

[35] A. A. Pakhnevich, S. V. Golod, and V. Y. Prinz, "Surface melting of copper during graphene growth by chemical vapour deposition," *J. Phys. D. Appl. Phys.*, vol. 48, no. 43, 2015.

[36] ASM International, *Casting design and performance*. ASM International, 2009.

[37] Y. Fan, K. He, H. Tan, S. Speller, and J. H. Warner, "Crack-Free Growth and Transfer of Continuous Monolayer Graphene Grown on Melted Copper," 2014.

[38] Lambert M. Surhone, Miriam T. Timpledon, and Susan F. Marseken, Eds., *Planck's Law: Black Body, Radiance, Electromagnetic Radiation, Wavelength*. Betascript Publishing, 2010.